\documentclass[11pt]{article}
\usepackage{amsmath,amssymb,graphicx}

\renewcommand{\title}[1]{
\begin{center} \Large \bf #1 \end{center}
}
\renewcommand{\author}[2]{
\begin{center} { #1}  \vspace{2mm}\\ 
  {\it #2}
\end{center}
\addvspace{\baselineskip}
}

\input{tcilatex}
\begin{document}

\title{Probing New Physics in Low Energy Solar Neutrino Oscillation Data}
\author{Amir N. Khan$^{a}$ (ntrnphysics@gmail.com), Douglas W. McKay$^{b}$
(dmckay@ku.edu)}

\setcounter{section}{0} \setcounter{equation}{0} \setcounter{figure}{0} %
\baselineskip5mm 

%

$^{a}$School of Physics, Sun Yat-Sen University, Guangzhou 510275, P. R.
China{\ }

$^{b}$Department of Physics and Astronomy, University of Kansas, Lawrence,
KS 66045, USA%

\textbf{Introduction}\footnote{\textit{Based on the talk presented by ANK at
the 6th CST-MISC Joint Symposium on Particle Physics -- from Spacetime
Dynamics to Phenomenology --, Oct. 15-16, 2016, Campus Plaza Kyoto, Kyoto,
Japan.}
\par
\bigskip}

Solar neutrinos have a long history since its first detection in 1970 in the
Homestake experiment, the epic of solar neutrino problem, which was an
apparent discrepancy between the Homestake observation and the theory based
on the solar standard model. This discrepancy or the solar neutrino problem
was confirmed in 1980s by Kamiokande, GNO/GALLEX, SAGE and Super-Kamiokande
(1990s) \cite{PDG}. The solar neutrino problem was finally solved by SNO
experiment in 2002 using the $^{8}$B component of the solar neutrino
spectrum and was shown consistent with the expectations of the solar
standard model in addition to the theory of neutrino oscillations \cite{SNO}.

Solar neutrinos have some unique characteristics. They cover the energy
range from sub-MeV to several MeVs. Having low energies they require
difficult detection techniques. They play complimentary role to the reactor
neutrinos. Both have approximately the same energy window for their
detections (0-10) MeV. In case of solar neutrinos, the neutrino flavor is
produced as result of the thermonuclear conversion of proton into neutrons
and positrons, while electron-antineutrinos are produced as a result of
neutron beta decay along with electrons. In both cases the neutrinos are
detected in the disappearance experiments using the similar detection
techniques.

Before the Borexino experiment, the solar neutrino fluxes were only known
indirectly by the radiochemical experiments. It was only very recently when
the Borexino experiment gave the first real-time flux observations of the
lowest energy components of the solar neutrinos (pp \cite{Borexino1}, $^{7}$%
Be \cite{Borexino2} and pep \cite{Borexino3}). Being at the lower end of the
solar neutrino spectrum, these components are predominantly controlled by
the vacuum oscillations. The LMA-MSW contribution to the three components
are \TEXTsymbol{<}2\% for pp, \TEXTsymbol{<}4\% for $^{7}$Be and \TEXTsymbol{%
<}8\% for pep. Based on this observation the recent solar neutrino flux
observations by the Borexino experiment are ideal to probe for nonstandard
neutrino interactions (NSIs) at the sun and at the detector while the NSIs
at the propagation inside the sun can be safely ignored because of the low
contribution from the LMA-MSW effect to pp, $^{7}$Be and pep components of
the spectrum.

\textbf{Calculational Framework:}

We estimate the source and detector NSI parameters in terms of the neutrinos
detection event rates at the Borexino detector which we define as%
\begin{equation}
R_{\nu }=N_{e}\int_{0}^{E_{max}}dE_{\nu }\phi (E_{\nu })\left( \sigma
_{e}(E_{\nu })\langle P\rangle _{ee}+\sigma _{\mu ,\tau }(E_{\nu
})[1-\langle P\rangle _{ee}]\right) ,
\end{equation}

where $N_{e}$ are the number of target electrons at the Borexino detector, $%
\phi (E_{\nu })$ is neutrino flux for the three pp,$^{7}$Be and pep taken
from Refs. \cite{Borexino1}, \cite{Borexino2} and \cite{Borexino3},
respectively. $\sigma _{e}(E_{\nu })$ and $\sigma _{\mu ,\tau }(E_{\nu })$
are the total $\nu _{e}$-e and $\nu _{\mu /\tau }$-e cross-sections and $%
\langle P\rangle _{ee}$ is the probability averaged over the oscillation
length that $\nu _{e}$ survives in the trip from the Sun's core to the
detector. $\langle P\rangle _{ee}$ contains all of the standard mixing
parameters and all of the source NSI\ parameters while $\sigma _{e}(E_{\nu
}) $ and $\sigma _{\mu ,\tau }(E_{\nu })$ contains the standard weak mixing
angle as well as all of the detector NSI\ parameters. For the statistical
estimates of standard weak mixing angle and all of the source and detector
NSI parameters we use the following $\chi ^{2}-$model

\begin{equation}
\chi ^{2}=\underset{i}{\dsum }\left( \frac{R_{i}^{\text{exp}}-R_{i}^{\text{%
obs}}}{\Delta _{i}^{\text{stat}}}\right) ^{2},
\end{equation}%
where $R_{i}^{\text{exp}}$are expected event rates as given in Eq. (1), $%
R_{i}^{\text{obs}}$ are the observed event rates as given in Refs. \cite%
{Borexino1}, \cite{Borexino2} and \cite{Borexino3} with $\Delta _{i}^{\text{%
stat}}$ as the statistical uncertainty. Index "$i$" designates the three
fluxes corresponding to pp, $^{7}$Be and pep.

\textbf{Analysis Details, Results and Discussion:}

For the SM fit, we set all NSI parameters to zero and fit the weak mixing
angle to Borexino data of pp, $^{7}$Be and pep spectra. For all\ the NSI
parameter fits, we perform two parameters space analyses, while set all the
other parameters to zero, in the order source-only, detector-only and then
source vs. detector. For the whole NSIs analyses, we assume the PDG-14
values of all the standard parameters to calculate the expected event rates.
All the two parameter regions are taken at 68\%, 90\% and 95\% C.L.. The
bounds are extracted in all of the above three cases at the 90\%C.L..

In case of the standard mixing model parameter analysis we find the value of
sin$^{2}\theta _{W}$ at the lowest energy to-date. From the fitting with the
source-only, detector-only and source vs. detector only NSI parameters we
find the bounds comparable to the existing bounds of Ref. \cite{NUNU} in
some cases and stronger or weaker in the other cases. The 90\% C.L.
boundaries of the flavor conserving nonuniversal NSIs parameters (left
panel) and the flavor-changing parameters (right panel) for the new physics
effects at detector for the Borexino's data for pp, $^{7}$Be and pep spectra
are shown in Fig. 1 for illustration. We also find the future prospects
estimates of the confidence level boundaries by incorporating 1\% of the
statistical uncertainty in regards to the future experiments for the solar
flux measurements at its lower energy end at Borexino (upgraded), CJPL \cite%
{CJPL}, SNO+ \cite{SNOP}, LENA \cite{LENA}, JUNO \cite{JUNO} etc. With 1\%
uncertainty we find an over all of one order of magnitude or more
improvement of the source and detector NSI parameters. Moreover, we also
check the impacts of the NSI phases on the C.L. boundaries of various NSI
parameters. The 90\% C.L. boundaries of the flavor conserving nonuniversal
NSI parameters (left panel) and the flavor-changing parameters (right panel)
for the new physics effects at the detector with the prospected data for pp, 
$^{7}$Be and pep spectra with 1\% uncertainty are shown in Fig. 2 for
illustration. 
\begin{figure}[tbph]
\begin{center}
\includegraphics[width=5in]{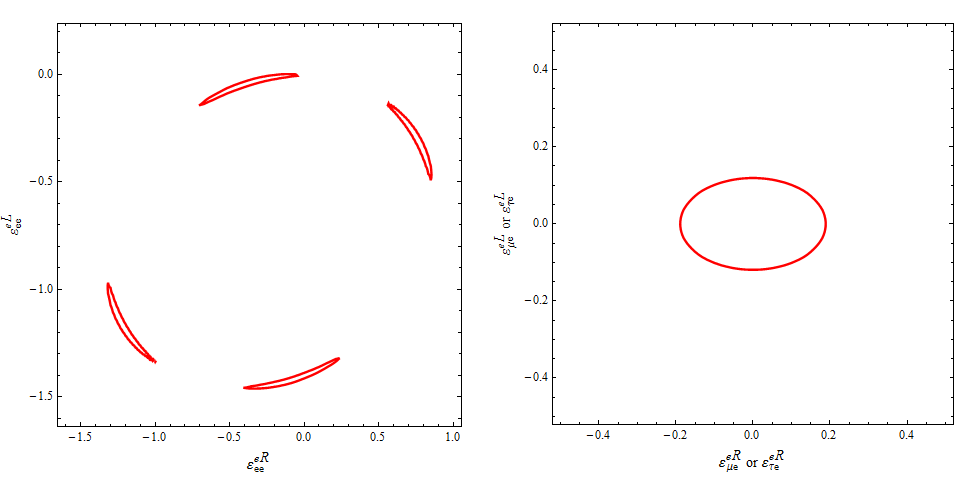}
\end{center}
\par
.
\caption{The 90\% C.L. boundaries of the flavor conserving nonuniversal NSIs
parameter (right panel) and the flavor-changing parameters(left panel) at
detector for the Borexino's data for pp, $^{7}$Be and pep spectra.}
\label{fig:1&2paramfits}
\end{figure}

\begin{figure}[tbph]
\begin{center}
\includegraphics[width=5in]{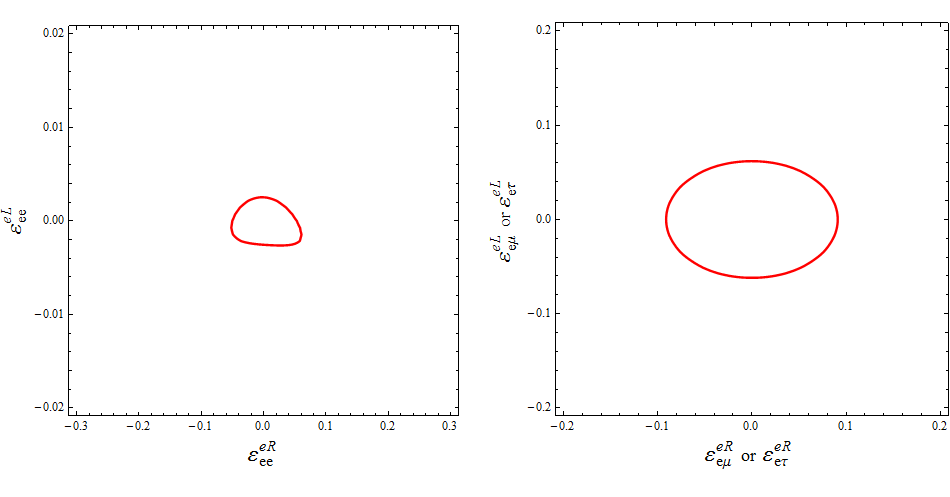}
\end{center}
\par
.
\caption{The 90\% C.L. boundaries of the flavor conserving nonuniversal NSIs
parameter (left panel) and the flavor-changing parameters (right panel) at
detector for the prospected data for pp, $^{7}$Be and pep spectra at the
future facilities.}
\end{figure}

\textbf{Conclusions:}\textit{\ }The recent real-time measurements of low
energy components of the solar spectrum at Borexino have very low LMA-MSW
contribution, thus provide a good testing ground for new physics study at
source (Sun) and detector. We found the best fit value of $\sin ^{2}\theta
_{W\text{ }}$at the lowest energy to-date using the Borexino results. We
have put new constraints on the NSI parameters at the production point at
Sun and at detector using the current data and have future prediction study
for the future proposals/planned experiments Borexino (upgrade) \cite%
{Borexino1}, CJPL \cite{CJPL}, SNO+ \cite{SNOP}, LENA \cite{LENA}, JUNO \cite%
{JUNO} etc. An improvement in sensitivity to the 1\% level will either
reveal very small deviations from the SM or reduce possibilities for NSI
parameters by factors from 2-3 to more than an order of magnitude. Our
results show the complementarity between solar and reactor data to probes
NSIs simultaneously. As a crucial background in the dark matter experiments,
solar neutrino experiments and theory can anticipate a long future. In
return, they provide the key to nailing down details of solar structure and
dynamics and can play a vital part of progress in resolving the neutrino
properties.

\textbf{Acknowledgement: }A. N. Khan\textbf{\ }is\textbf{\ }very thankful to
Professor Joe Sato and Professor Osamu Yasuda who supported his visit and
stay at Saitama University, Japan. He also thanks to the organizers of the
"6th CST-MISC Joint Symposium on Particle Physics

-- from Spacetime Dynamics to Phenomenology --". at Maskawa Institute,
Kyoto-Sangyo University, for their kind hospitality during his stay there
and giving him the opportunity to present this work. This work was
financially supported by the Sun Yat-Sen University, China, under the
Post-Doctoral Fellowship program.



\end{document}